\documentclass[reprint,groupedaddress,amsmath,amssymb,pra,noeprint]{revtex4-2}
\usepackage[utf8]{inputenc}
\usepackage{amsmath}
\usepackage{graphicx}
\usepackage{siunitx}
\usepackage{color}
\usepackage{upgreek}
\usepackage{comment}
\usepackage[normalem]{ulem}
\usepackage{setspace}
\usepackage{braket}

\begin{document}

\title{Optimisation of Scalable Ion-Cavity Interfaces for Quantum Photonic Networks}

\author{Shaobo Gao}
\affiliation{Department of Physics, University of Oxford, Clarendon Laboratory, Parks Road, Oxford OX1 3PU, United Kingdom}
\author{Jacob A Blackmore}
\email{Corresponding Author: jacob.blackmore@physics.ox.ac.uk}
\affiliation{Department of Physics, University of Oxford, Clarendon Laboratory, Parks Road, Oxford OX1 3PU, United Kingdom}
\author{William J Hughes}
\affiliation{Department of Physics, University of Oxford, Clarendon Laboratory, Parks Road, Oxford OX1 3PU, United Kingdom}
\author{Thomas H Doherty}
\affiliation{Department of Physics, University of Oxford, Clarendon Laboratory, Parks Road, Oxford OX1 3PU, United Kingdom}
\author{Joseph F Goodwin}
\email{joseph.goodwin@physics.ox.ac.uk}
\affiliation{Department of Physics, University of Oxford, Clarendon Laboratory, Parks Road, Oxford OX1 3PU, United Kingdom}

\date{\today}

\begin{abstract}
In the design optimisation of ion-cavity interfaces for quantum networking applications, difficulties occur due to the many competing figures of merit and highly interdependent design constraints, many of which present `soft-limits', amenable to improvement at the cost of engineering time. In this work we present a systematic approach to this problem which offers a means to identify efficient and robust operating regimes, and to elucidate the trade-offs involved in the design process, allowing engineering efforts to be focused on the most sensitive and critical parameters. We show that in many relevant cases it is possible to approximately separate the geometric aspects of the cooperativity from those associated with the atomic system and the mirror surfaces themselves, greatly simplifying the optimisation procedure. Although our approach to optimisation can be applied to most operating regimes, here we consider cavities suitable for typical ion trapping experiments, and with substantial transverse misalignment of the mirrors. We find that cavities with mirror misalignments of many micrometres can still offer very high photon extraction efficiencies, offering an appealing route to the scalable production of ion-cavity interfaces for large scale quantum networks.
\end{abstract}

\maketitle

\section{Introduction}\label{sec:intro}

There are many situations in which the interaction of atoms and light may be usefully augmented by placing the atomic system within an optical cavity, ranging from highly efficient optical lattice generation~\cite{Hamilton2015,Park2021} to the creation of spin squeezed states for improved metrology~\cite{Braverman2019,PedrozoPenafiel2020,Muniz2021}, as well as a wealth of quantum information processing applications~\cite{vanEnk2004,Duan2004,Cho2005,Munro2012,Reiserer2014}. Among these, one application where cavities seem poised to play a critical role is in the efficient extraction of single photons from trapped atoms or ions. Such systems will likely be central to matter-light interface design in quantum networks~\cite{Kuhn2002,Moehring2007,Ritter2012,Northup2014,Kobel2021} and, at a more basic level, can also be used as extremely high quality sources of indistinguishable single photons~\cite{Kuhn2002,Keller2004,Maurer2004,McKeever2004,Hijlkema2007,Barros2009,sterk2012,Muecke2013,Walker2020}.

Despite a decade of significant advances~\cite{kimble2008,Reiserer2015,takahashi2020,Schupp2021} in the use of cavities for network applications, the plethora of varied designs seen in today’s experiments~\cite{Podoliak2016} suggests an interesting question: Given a few simple experimental constraints, is it possible to identify an optimal atom-cavity system for quantum networking, and what defines `optimal’, given the many competing figures of merit? Fidelity, attempt rate and success probability are clearly important considerations, but for practical systems simplicity of construction and stability are also paramount. Even if a figure of merit can be identified, the problem is complicated by the many different design parameters and a variety of often experiment-specific parameter constraints and manufacturing tolerances.  It is to our benefit that few of these constraints are absolute, but such ‘soft-limits’ make simple many-parameter optimisation a poor substitute for a real understanding of where the easiest routes to improvement lie.

Identifying the dominant drivers of performance in atom-cavity systems is crucial if such devices are to be produced at scale for the realisation of quantum networks of useful size. The requirement that a large number of such devices work simultaneously requires each to be extremely reliable, a phrase rarely heard in relation to today's cavity networking experiments. Achieving this means on the one hand identifying parameter regimes where robust performance may be realised; and on the other minimising the experimental and engineering complexity of the systems used.

In this manuscript we set out a systematic approach to addressing the first of these challenges. We focus particularly on the use of such systems in the rapid generation of entanglement between remote pairs of atomic ions via a measurement-based entanglement-swapping scheme, a key resource for distributed quantum computation~\cite{Monroe2013,Nickerson2014,Stephenson2019}. While many of the ideas and approaches presented may by readily applied to other use-cases and types of quantum emitter (e.g. neutral atoms), care should be taken to ensure the assumptions made still apply. Most notably, many trapped atom applications require very low-linewidth cavities to ensure sufficient resonant selection of the desired atomic transition, meaning it may not be possible to optimise the cavity geometry independently of the mirror coating, as we will in this work.

We start in Sec~\ref{sec:def} by providing a detailed definition of the problem, describing the photon production scheme, cavity geometry and `real-world’ limits and tolerances of the system. In Sec~\ref{sec:metrics} we then consider performance metrics for ion-cavity systems in the context of quantum networking. Despite the complexity of the general case, we show that with appropriate choice of atomic levels and driving scheme, a much simpler performance optimisation is sufficient, namely maximising the probability of extracting a single photon from the cavity. In Sec~\ref{sec:optimise} we show that the optimisation of extraction probability is well approximated as a separable problem, with the atomic, geometric and mirror interface parameters considered in turn. We then show how this can be used to identify optimal geometries and mirror coatings given limits on mirror misalignment or scattering loss. In Sec~\ref{sec:perform} we use this separability to study the impact of realistic constraints on the optimisation and how the performance of the cavity is impacted by errors in manufacture or in estimation of the magnitude of losses and misalignment. We conclude in Sec~\ref{sec:outlook} with a summary of the operating regimes where performance is achievable with the lowest engineering demands. Most notably, we find that (with some caveats) substantial transverse misalignments of the cavity mirrors are indeed tolerable, underlining the potential for simple, passively aligned cavity assemblies suitable for scalable quantum networks.

\section{Description of the Problem}\label{sec:def}

\subsection{Photon Production Scheme}\label{ssec:driving}

\begin{figure}[t]
    \centering
    \includegraphics[width=0.66\columnwidth]{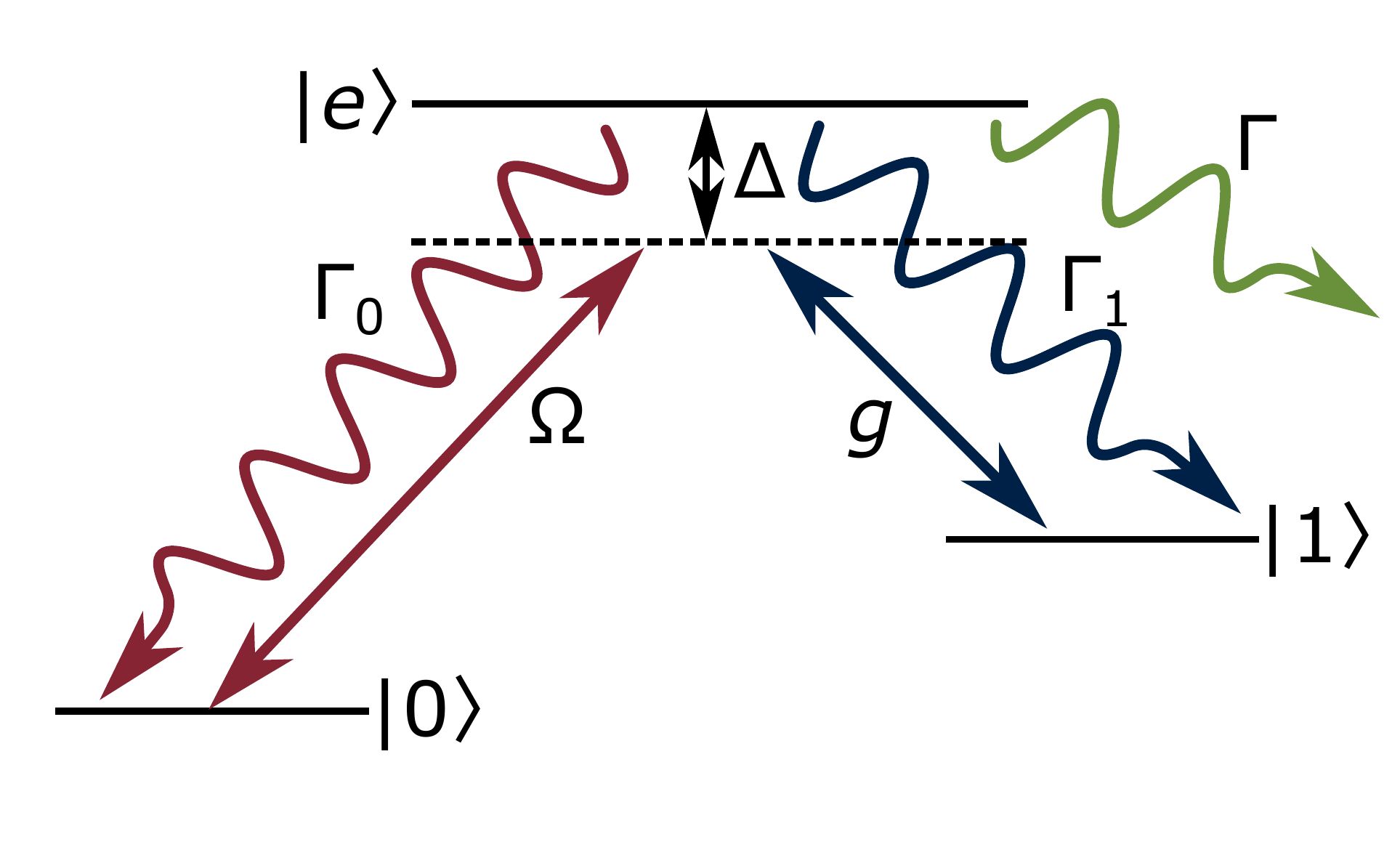}
    \caption{Simplified energy level structure for the atomic part of our system. A pair of (meta)stable energy levels, $\ket{0}$ and $\ket{1}$, are connected by a pair of dipole-allowed optical frequency transitions via a common excited state $\ket{e}$. The $\ket{0}\rightarrow\ket{e}$ transition is driven by an external laser field, with Rabi frequency $\Omega$. The transition $\ket{e}\rightarrow\ket{1}$ is driven by the vacuum mode of the cavity, with ion-cavity coupling strength $g$. The laser and cavity fields may be tuned away from resonance with the bare atomic level by an equal amount $\Delta$ to reduce population of $\ket{e}$. From $\ket{e}$ we assume that decays are possible to $\ket{0}$ with rate $\Gamma_0$, to $\ket{1}$ with rate $\Gamma_1$, and to possible other levels (not shown) to give a total upper state decay rate of $\Gamma$. Transfer of population to $\ket{1}$ is associated with emission of a single photon of frequency $\omega_{e1}-\Delta$, which must occur into the cavity mode to be useful for networking.}
    \label{fig:atom system}
\end{figure}

There are several possible ways to extract a photon from an atomic system into an optical cavity mode \cite{Bochmann2008, Barros2009}. We choose to assume use of a vacuum-mediated stimulated Raman adiabatic passage (vSTIRAP) scheme~\cite{Kuhn2002,Vasilev2010,Shore2017,Bergmann_2019}, a simplified illustration of the level structure of which is given in Fig~\ref{fig:atom system}. In this scheme, the ion is initialised in level $\ket{0}$ and transferred to state $\ket{1}$ via emission of a single photon near the $\ket{e}\rightarrow\ket{1}$ transition frequency. States $\ket{0}$ and $\ket{1}$ are generally stable or metastable and not directly connected by an electric dipole (E1) transition, instead each being connected by E1 transitions to an excited state $\ket{e}$, creating a $\Lambda$-type system. The transition $\ket{0}\rightarrow\ket{e}$ is driven by an external laser source with Rabi frequency $\Omega$; we label this arm of the transition the `drive'. The other arm is the `cavity' transition from $\ket{e}\rightarrow\ket{1}$, driven by the cavity vacuum field with strength quantified via the ion-cavity coupling $g$. For efficient transfer both arms of the transition must have equal detuning $\Delta$ such that the Raman condition is met~\cite{Bergmann_2019}. This family of schemes has many advantages over simple two-level {$\pi$--pulse} schemes: control of the photon wavepacket shape~\cite{Keller2004,Nisbet_Jones2011}; compatibility with almost any dipole-allowed $\Lambda$-system in the atom, enabling a wide range of network photon wavelengths; compatibility with bichromatic driving schemes~\cite{Krutyanskiy2019}; and maximisation of photon extraction probability, provided one is unconcerned with the length of the photon (see discussion of this point in Sec.~\ref{ssec:rate} and App.~\ref{app:Adiabatic}).

Measurement-based remote entanglement schemes must first produce a bipartite entanglement between an atomic qubit and a photonic qubit. To encode and populate these qubits we will require that $\ket{1}$ represents a manifold of several distinct sublevels; typically this will also be the case for $\ket{0}$ and $\ket{e}$. Many protocols are possible \cite{Luo2009}, but each involves driving two discrete $\Lambda$ systems. These may or may not share an initial state, and may be driven simultaneously with a bichromatic drive~\cite{Stute2012,Stephenson2019,Krutyanskiy2019} (for polarisation-encoded schemes) or individually in sequence~\cite{Nisbet_Jones2013} (for time-bin encoded schemes).

The presence of the more complex electronic structure necessary to produce spin-photon entanglement introduces extra conditions which must be met to avoid fidelity loss due to off-resonant transfers and/or spectator decay outside the qubit subspace~\cite{gao2020} (see Sec~\ref{sec:metrics}). However, providing these are satisfied, the performance of each of the pair of $\Lambda$ systems is well-described by the model given in Fig~\ref{fig:atom system}. For the purposes of this paper we optimise this simple system only, with the results applicable -- with appropriate caveats -- to a diverse range of remote-entanglement schemes.

\subsection{The Cavity}\label{ssec:cavity}
\begin{figure}[t]
    \centering
    \includegraphics[width=0.9\columnwidth]{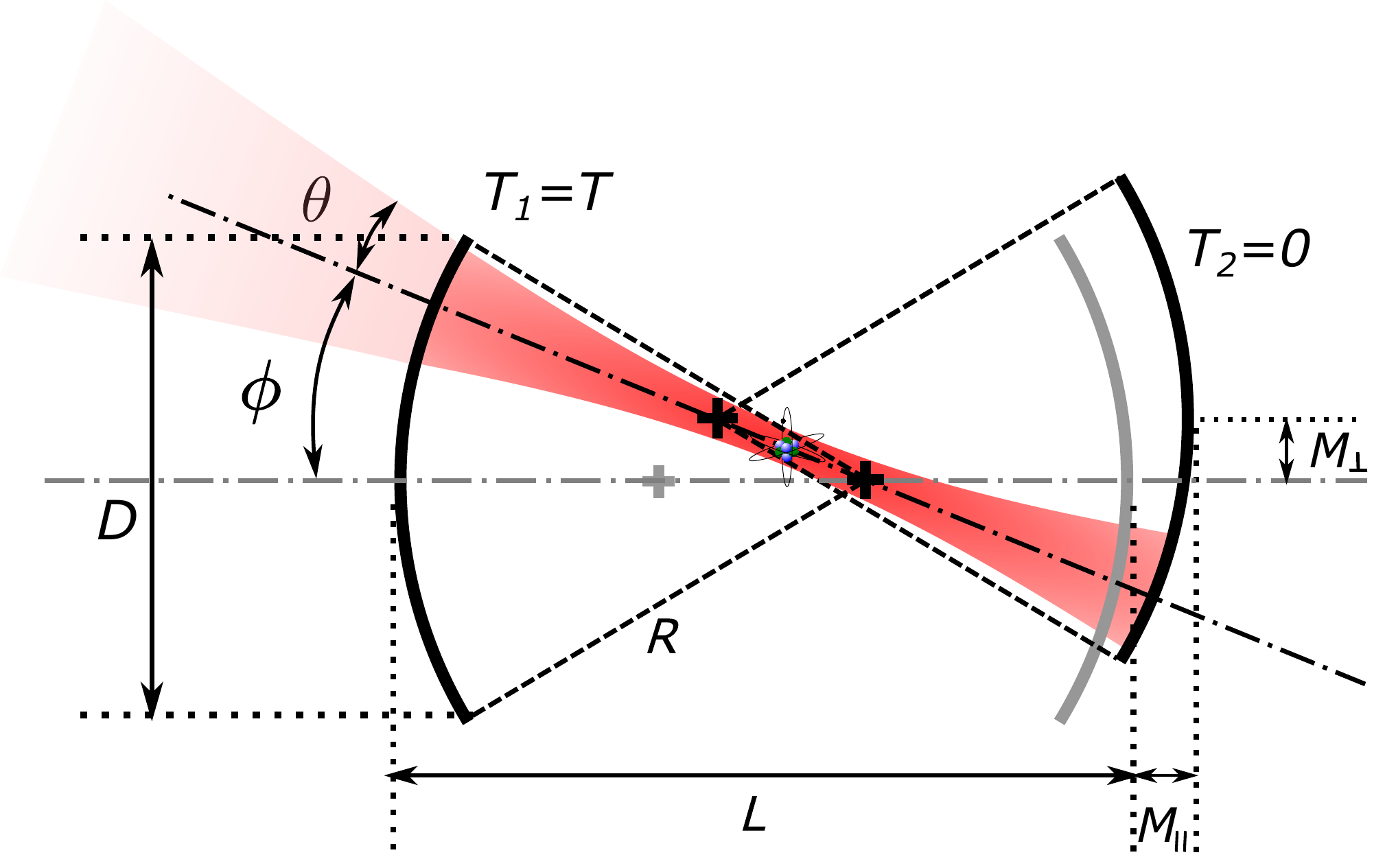}
    \caption{The geometry describing a misaligned optical cavity. The cavity consists of two concave spherical mirrors with equal radii of curvature ($R$) and supports a mode with divergence half-angle $\theta$. In the optimum configuration (shown in grey) the overall length of the cavity is $L$ and the two mirrors are perfectly aligned to the cavity axis. In the misaligned case one mirror is displaced along the cavity axis by $M_\parallel$ and transversely by $M_\perp$; we assume for simplicity that $M_\parallel=M_\perp=M$. This misalignment tilts the overall mode by an angle $\phi$ from the cavity axis and slightly increases $\theta$.}
    \label{fig:cavity system}
\end{figure}

Having defined the photon production scheme within the electronic structure of the emitter, we now describe the properties of the Fabry-P\'erot cavity used to provide the coupling, $g$. We consider the symmetric cavity geometry shown in Fig~\ref{fig:cavity system}, consisting of two concave spherical mirrors with equal radii of curvature $R$ and equal diameter $D$. In the ideal system the mirrors are co-axial and separated by $L$, and the ion sits on the cavity axis at $L/2$. The volume of the mirrors, modelled as a spherical cap is given by
\begin{equation}
    V = \frac{\pi h}{6} \left(\frac{3D^2}{4}+h^2 \right),
\end{equation}
where the saggita, $h$, is defined as $R-\sqrt{R^2-D^2/4}$. Within this framework the fundamental lower limit to $L$ is $2h$, corresponding to the two mirrors touching when not misaligned. 

We consider coupling to the fundamental transverse electric $\rm{TE}_{00}$ mode of the cavity and neglect higher-order components, which can be ensured by tuning the fundamental mode to Raman resonance and choosing a suitable $L/R$ ratio to break the transverse mode degeneracy. Our mode then forms a TE\textsubscript{00} Gaussian mode with waist $w_0$ at the location of the ion, with a far-field divergence half-angle $\theta$, where
\begin{subequations}
\begin{equation}
    \theta = \frac{\lambda}{\pi w_0},
    \label{eq:theta}
\end{equation}
\begin{equation}
    w_0 = \sqrt{\frac{\lambda}{2\pi}}\bigg[L(2R-L)\bigg]^{1/4},
    \label{eq:w0}
\end{equation}
\end{subequations}
and $\lambda$ is the wavelength of the mode. At the centre of the cavity, an emitter located at an antinode of the cavity mode couples to the cavity with a rate $g_0$ inversely proportional to the square root of the mode volume $V=\pi w_0^2 L/4$:
\begin{equation}
    g_0=\sqrt{\frac{3\lambda^2 c \gamma_1}{\pi^2 L w_0^2}}=\sqrt{\frac{3\lambda^2 c \gamma_1}{4\pi V}},
    \label{eq:g0}
\end{equation}
where $c$ is the speed of light in vacuum and $\gamma_1=\Gamma_1/2$ is the partial (half-)linewidth of the atomic decay channel coupled to the cavity. Note that if the vSTIRAP transfer is between sublevels of the initial, final and excited states, this represents the partial linewidth of the specific transition, including angular momentum considerations. Furthermore, we assume that the cavity and magnetic field are arranged such that the cavity vacuum mode couples optimally to the transition in question; where this is not the case~\footnote{E.g. in the case of a cavity with optical axis near-parallel to the magnetic field, driving a $\pi$ transition}, $g_0$ will be suppressed accordingly.

Residual translational misalignment of the cavity mirrors can occur in the transverse and longitudinal directions, with magnitudes $M_\perp$ and $M_\parallel$ respectively. For cavity configurations far from concentricity, $M_\parallel$ has negligible effect, while $M_\perp$ primarily introduces a tilt $\phi$ to the cavity mode axis, measured relative to the unperturbed value. However, for near--concentric cavities, $M_\parallel$ significantly changes the sensitivity of the system to transverse misalignment, and furthermore leads to substantial changes in $\theta$:
\begin{subequations}
\begin{equation}
    \phi=\arctan{\left(\frac{M_\perp}{[2R-L]-M_\parallel}\right)}\approx\frac{M_\perp}{[2R-L]-M_\parallel},
\end{equation}
\begin{equation}
   L\rightarrow L' = 2R-\left(\frac{2R-L-M_{\parallel}}{\cos{\phi}}\right)\approx L+M_\parallel,
    \label{eq:actual length} 
\end{equation}
\begin{equation}
    \theta \rightarrow \theta' =\left[\frac{2\lambda}{\pi\sqrt{L'(2R-L')}}\right]^{1/2}.
    \label{eq:thetaprime}
\end{equation}
\end{subequations}
These distinct effects mean the two forms of misalignment should in general be considered independently;
to simplify our optimisation we choose to define a single misalignment parameter $M=M_\parallel=M_\perp$. This combination of the two parameters represents a worst-case scenario, where increasing axial misalignment moves the cavity closer to concentricity, increasing the sensitivity to transverse misalignment. This is itself compounded by radial misalignments increasing $\phi$. The action of both is to increase the clipping losses, as shown in Fig.~~\ref{fig:clipping vs L R D M}. Note that for the purposes of this work we assume that the ion position is appropriately adjusted to remain aligned to the mode centre when this shifts due to mirror misalignment.

To ensure photons are extracted from the cavity into a single free-space mode, one mirror is coated for maximum reflection ($T_2\approx0$), while the other `outcoupler' mirror has a finite transmission $T_1=T$. The value of $T$ will be chosen to optimise the performance of the system (see~\ref{ssec:scattering}) and is typically several hundred parts per million. Considering these transmissive losses along with the \emph{intrinsic} round-trip losses $\mathcal{L}_\text{in}$ due to absorption, scattering or clipping of the cavity mode; the total decay rate of the cavity field amplitude, $\kappa$; and finesse $\mathcal{F}$ are thus given by:
\begin{subequations}
\begin{equation}
    \kappa=\frac{c}{4L}\left(T+\mathcal{L}_\text{in}\right),
\end{equation}
\begin{equation}
    \mathcal{F}=\frac{2\pi}{T+\mathcal{L}_\text{in}}.
\end{equation}
\end{subequations}

We consider two distinct contributions to $\mathcal{L}_\text{in}$, assuming that other losses, such as absorption or scattering in the region between the cavity mirrors, are negligible for short cavities under ultra-high vacuum conditions:
\begin{enumerate}
    \item{$\mathcal{L}_\text{clip}$ describes losses due to clipping of the edge of the mode by the finite diameter of the cavity mirror~\footnote{Hard clipping by an aperture located on the cavity axis away from the mirrors will have a similar effect, with dependence on diameter scaled according to the expected mode diameter at the aperture location}}. 
    \item $\mathcal{L}_\text{scat}$ describes scattering losses due to RMS mirror surface roughness ($\sigma$), typically approximated \cite{Bennett1992} by
\begin{equation}\label{eq:loss}
\mathcal{L}_\text{scat}\approx 2\left[1-e^{ - (4\pi\sigma / \lambda)^2}\right].
\end{equation}
\end{enumerate}
As we will show in Sec~\ref{sec:optimise}, the assumption that these two loss mechanisms are independent underpins our approach to optimisation.

\subsection{Practical Considerations and Assumptions}\label{ssec:practical}

Although the methods described in this paper are reasonably general and may easily be adapted to most emitter-cavity systems, we make a number of assumptions regarding the geometry of the cavity and mirrors motivated by our own experiments, which we outline here.

The cavity is assumed to be arranged around the centre of an ion trap, and as such is limited in its minimum length and maximum mirror diameter, due to the need to minimise exposure of the ion to the mirror dielectric, which can disrupt the trapping potential~\cite{Podoliak2016}. The ion is assumed to be located at the centre of the cavity mode, and perfectly localised.

The mirrors are assumed to exhibit significant relative transverse misalignment -- although the treatment holds for more precisely aligned mirrors, the range of $M$ we consider is more representative of the offsets which might occur in simple, passively aligned systems. Meanwhile, angular misalignment of the mirrors is considered to be insignificant and the length of the cavity is assumed to be stable and set at an appropriate value for the necessary cavity-Raman resonance, achieved in our system by active closed-loop control of this degree of freedom.

The mirrors themselves are assumed for simplicity to be perfectly spherical, with uniform scattering loss across the surface. This is a good approximation for some fabrication methods (mechanical or focused ion beam milling), but traditional laser ablation methods typically yield Gaussian mirror profiles, and even modern dot-milling ablation techniques lead to significant deviations from a spherical profile away from the mirror centre~\cite{Ott:16}. Empirically, however, `clipping'-type losses still grow rapidly with increasing $\phi$ or $\theta$ and to a suitable approximation (see Eq.~\ref{eq:clipping_loss} and associated discussion) Gaussian and other non--spherical mirrors can be well--modelled by a hard--edged spherical profile~\cite{Benedikter_2015}.

For the purposes of the analysis presented in this article we consider the machining of mirrors to be volume limited i.e. there is a cost associated with removing a certain volume of the mirror substrate, be it in one or more of milling time, equipment limitations, surface quality degredation \textit{et cetera}. Bounded by this limit we assume that during the manufacture one can freely vary $R$ and $D$ to maximise performance.

\section{Performance Metrics}\label{sec:metrics}

The overriding aim of a cavity-mediated quantum network is to produce entanglement between pairs of remote nodes at high fidelity and rate. Currently, the state-of-the-art fidelity is for ion-ion remote entanglement at 96.0(1)\%, which was achieved at a rate of $\SI{100}{\second^{-1}}$, collecting photons with a high-numerical-aperture lens~\cite{Nadlinger2022}. 
Systems for practical future quantum networks will require substantially higher rates, and similar or higher fidelities, to ensure entanglement between nodes can be efficiently distilled to fidelities approaching those of local gates~\cite{Nigmatullin2016}.
However, how these two key metrics should be measured against one another is a non-trivial question and is highly architecture- and application-dependent. Furthermore, many factors contribute to the performance, including purely atomic processes such as state preparation and mapping, and purely photonic considerations such as imperfections in the photonic Bell-state analyser. Given these concerns, it is reasonable to question whether a universal approach to network cavity design optimisation exists. 

In this section, we will consider the impact on remote entanglement fidelity and rate due to properties of the ion-cavity system itself, and demonstrate that for suitable choices of atomic system and driving scheme it is possible to identify a single metric of performance, independent of the finer details of the network architecture.

\subsection{Fidelity}\label{ssec:fidelity}

We aim to achieve high fidelity entanglement between pairs of ions located in remote nodes via a two-photon entanglement-swapping protocol~\cite{Moehring2007,Stephenson2019}. This requires the production of single photons from each node, with one photonic degree of freedom (e.g. polarisation or time-bin) entangled with high fidelity to the final electronic state of the atom. In order for the path information of the two photons to be erased and the entanglement swapped from ion-photon to ion-ion, each photon must also be highly indistinguishable. We now consider three mechanisms which can reduce the fidelity of ion-photon entanglement or the indistinguishability of the photons produced.

First, we note that the fidelity of the ion-photon state may be reduced due to off-resonant coupling during the vSTIRAP transfer. As discussed in Sec~\ref{ssec:driving}, transfer to an (entangled) electronic qubit superposition state requires two $\Lambda$-systems to be driven, and unless these two processes are made sufficiently distinct via frequency and/or polarisation selection, the ion and photon will only be partially entangled. Additionally, off-resonant decays to other sublevels beyond the electronic qubit subspace introduce problematic `loss' errors; for more details see~\cite{gao2020, DohertyThesis}. Avoiding these effects generally demands that the qubit levels are well-resolved, i.e. the final qubit splitting is much larger than $\kappa$. This is an important consideration but such errors can typically be made negligible via suitable choice of atomic system and bias field.

To produce indistinguishable photons, each node must emit well-defined and identical wavepackets. Interestingly, provided the photon detectors are time-resolving and with sufficiently low jitter, it is not necessary for the photons to be at identical frequencies \cite{Vittorini2014} but only for any difference to be stable and well-characterised. However, any mismatch in the wavepacket envelopes between nodes provides which-way information during the entanglement-swapping step, reducing the final fidelity. The use of vSTIRAP schemes permits the shape of the wavepacket envelope to be tuned via suitable modulation of the drive pulse, allowing such effects to be suppressed. A more significant problem occurs in the event of spontaneous scattering back to the initial state during the Raman transfer, the stochastic nature of which leaves the photonic wavepacket in a mixed state. We give an overview of this problem in Appendix~\ref{app:Mixing}, with detailed discussions also found in~\cite{Walker2020,gao2020}. While vSTIRAP pulse shaping can help reduce such spontaneous scattering, it is also crucial to bias the system towards transfer into the target state. One route to achieving this is to use a very high ion-cavity coupling $g$, but this places unnecessary demands on the precision of mirror alignment. A better approach is to simply use a $\Lambda$-system with a low branching ratio ($\alpha$) on the `drive' arm, which keeps fidelity losses due to spontaneous scattering $<1\%$ across the parameter ranges considered here~\cite{gao2020}, and often far lower, permitting us to conclude our consideration of this issue. 

Finally, we note that birefringence in the cavity mirror substrates or dielectric coatings can cause the polarisation of the intracavity field to rotate upon each reflection, introducing a time-dependence to the output photon polarisation. In polarisation-encoded schemes, although the fidelity of the ion-photon entanglement is only modestly reduced, the phase of this entangled state becomes time-dependent, which can significantly impact the achievable fidelity of the final ion-ion entangled state. We have recently studied the impact of such effects~\cite{Kassa2020}, concluding that without excessively complex mitigation strategies maintaining a fidelity loss $<1\%$ requires birefringence $\Omega_\mathrm{B}\ll\kappa/10$. This places considerable demands on the quality of mirrors used, and a more suitable approach may be to use time-bin encoded photonic qubits, a technique which also prevents fibre birefringence from limiting fidelities over longer link distances. For the purposes of this work we assume that the information is not encoded in the polarisation degree of freedom, and so the impact of any birefringence is negligible.

In summary, although these mechanisms should be considered carefully in the design of a photonic network, each is best mitigated through appropriate choice of atomic emitter, driving scheme, and photonic encoding. With these choices made, the impacts on fidelity due to the cavity itself are expected to be lower than those due to the photonic network, entanglement-swapping process and local operations on the atom, and we can consider only the maximisation of photon production rate in our optimisation process.

\subsection{Rate}\label{ssec:rate}

For a two-photon entanglement scheme~\cite{Moehring2007,Stephenson2019}, the rate at which we produce remote ion-ion Bell pairs is given by
\begin{equation}
    R_\text{Bell}=\frac{1}{2}\times\frac{(P_\text{ext}\epsilon_\text{net}\epsilon_\text{det})^2}{\tau_\text{att}},
\end{equation}
where $P_\text{ext}$ is the probability of single photon extraction from the cavity at each node, $\epsilon_\text{net}$ and $\epsilon_\text{det}$ are photon network transmission and detector efficiency, $\tau_\text{att}$ is the attempt cycle period, and the factor of $1/2$ accounts for anti-coincidence events and is intrinsic to the protocol~\cite{Moehring2007,Stephenson2019}. Note that due to the two-photon herald, the success rate is quadratically sensitive to the probability of single photon detection from each node. The network and detector efficiency terms depend on physical components of the system and lie beyond the scope of this paper, but we note for completeness that the wavelength of the network photon dictates the choice of materials and is thus a critical determinant of the performance. Optimising an ion-cavity system for high entanglement rate thus depends on maximising attempt rate $1/\tau_\text{att}$ and, most crucially, the photon extraction probability $P_\text{ext}$.

We first consider contributions to the attempt rate, considering contributions to the attempt cycle period:
\begin{equation}
    \tau_\text{att}\sim\tau_\text{prep}+\tau_\text{lat}+\tau_\gamma+\frac{\Delta x_\text{net}}{c}.
\end{equation}
Here, $\tau_\text{prep}$ is the time needed to optically pump and transfer population into the initial state, $\tau_\text{lat}$ is the sum of all optical and electronic latencies in the attempt loop, $\Delta x_\text{net}/c$ is the combined quantum and classical signal propagation time for two nodes separated by optical length $\Delta x_\text{net}$. The sum of these times limits the attempt rate to approximately \SI{1}{\mega\hertz}, and so by requiring the characteristic length of the photon $\tau_\gamma\ll\SI{1}{\micro\second}$ we can ensure the duration of photon production is insignificant. Most fundamentally, the rate at which a photon can be extracted from the cavity is limited by $\kappa$, but rapid production of a photon also demands a reasonable $g$ and a sufficiently strong Raman drive. However, driving the system too rapidly breaks the adiabaticity required for efficient population transfer, so some care must be taken in the choice of transfer pulse. In Appendix~\ref{app:Adiabatic}, we consider the effects of non-adiabatic driving, and show that for all the operating regimes outlined in this paper, high probability Raman transfer can be achieved while ensuring photon length makes an insignificant contribution to the attempt cycle period.

With appropriate care taken in the choice of atomic system, entanglement protocol and Raman transfer pulse, we have thus identified a single dominant figure of merit for our ion-cavity system, namely the probability of photon extraction per attempt cycle, $P_\text{ext}$.

To calculate $P_\mathrm{ext}$ we first consider the probability that a photon is generated in the cavity, $P_\mathrm{gen}$. Accounting for losses due to spontaneous emission we can place an upper bound on $P_\mathrm{gen}$, which saturates in the case of perfect adiabatic transfer:
\begin{subequations}
\begin{equation}
P_\text{gen}\leqslant\frac{2C}{2C+1},\label{eq:upper_bound}\\    
\end{equation}
\begin{equation}
C=\frac{g^2}{2\gamma\kappa}.    
\end{equation}
\end{subequations}
Here we have defined the cooperativity $C$, which is a function of $g$; and $2\gamma=\Gamma$, the total spontaneous decay rate of the excited atomic state. The overall extraction probability is then the product of $P_\text{gen}$ and the probability that the photon exits the outcoupler mirror before being absorbed or scattered out of the cavity mode:
\begin{equation}
    P_{\text{ext}}=P_\text{gen}\left(\frac{T}{T+\mathcal{L}_\text{in}}\right)\leqslant\frac{2C}{2C+1}\left(\frac{T}{T+\mathcal{L}_\text{in}}\right).
\end{equation}
Following~\cite{Goto2019}, $P_\text{ext}$ can be shown to be bounded by
\begin{equation}
    P_{\text{ext}} \leqslant 1-\frac{2}{1+\sqrt{1+2C_\mathrm{in}}}.
    \label{eq:probability}
\end{equation}
In which the \emph{intrinsic} cooperativity $C_\text{in}$ is defined as:
\begin{equation}
    C_{\text{in}} = \frac{g_0^2}{2\gamma\kappa_{\text{in}}},
    \label{eq:C_in1}
\end{equation}
where $g_0$ is the value of $g$ at the central cavity antinode. The intrinsic loss rate within the cavity is defined as
\begin{equation}\label{eq:lossin}
    \kappa_\text{in}= \frac{\mathcal{L}_\text{in} c}{4L}.
\end{equation}
The bound in Eq.\ref{eq:probability} is only saturated when the overall cavity decay, $\kappa$, is appropriately optimised by considering the optical properties of the cavity, as will be discussed further in Section~\ref{ssec:scattering}. As maximising finesse is not generally important, we assume that it is always possible to achieve this optimum $\kappa$, although this does require an accurate estimate of the intrinsic loss prior to coating the mirrors.

Note that in this treatment we have assumed that any photon populating the free-space mode leaking from the cavity outcoupler can be coupled into the photonic network with constant high efficiency. This is a reasonable assumption for a typical $\text{TE}_\text{00}$ cavity mode imaged appropriately into a single mode fibre, but for systems with large tilt $\phi$ and divergence $\theta$, this assumption may break down for certain cavity mirror substrate geometries.

\section{Cavity design optimisation}\label{sec:optimise}
In the following three subsections we will discuss how the geometric and optical properties of the cavity can be optimised to maximise photon extraction probability, for cavity geometries similar to those in Fig~\ref{fig:cavity system}. In Sec~\ref{ssec:optscheme} we discuss the parameters considered in the optimisation of the performance, and our strategy for optimisation. In Sec~\ref{ssec:clipping} we show how consideration of clipping losses defines the optimal cavity geometry and then use this result, along with the known mirror scattering loss, to optimise outcoupler transmission (and hence cavity finesse) in Sec~\ref{ssec:scattering}. In Sec~\ref{ssec:combined} we combine these results to understand how $P_\text{ext}$ is optimised through choice of $L$, $R$ and $T$, and offer some `rules of thumb' for identifying optimal operating regimes in the presence of certain practical constraints.

\subsection{Optimisation Scheme}~\label{ssec:optscheme}

In this subsection we will briefly outline our approach to design optimisation. In Sec~\ref{sec:metrics} we have shown, for certain common choices of driving and entanglement schemes, the optimisation of the ion-cavity system is reduced to a maximisation of the photon extraction probability
. As shown in Sec~\ref{ssec:rate}, $P_\mathrm{ext}$ is less than $P_\mathrm{gen}$ due to intrinsic losses in the cavity, and quantifying and minimising these losses will be central to our optimisation scheme.

The ability to split the sources of loss into independent constituents (see Eq~\ref{eq:loss}) enables the use of a two-part optimisation strategy, which we will describe in the following three sections. The first stage is to optimise the geometry by reducing the optimisation scheme to a maximisation of $g$ within well-defined bounds where $\mathcal{L}_\mathrm{clip}\ll\mathcal{L}_\mathrm{scat}$. The second stage is to then optimise $T$ given the fixed value of $\mathcal{L}_\mathrm{scat}$.

Parameters considered in the optimisation can be broadly split into two distinct categories. The first, which we shall call `controlled parameters', are those over which the system designer can tune to optimise performance. We consider the controlled parameters to be cavity length $L$, mirror radii of curvature $R$ and diameters $D$, and outcoupler transmission $T$. We assume the ability to set these parameter freely up to some limits defined by the experimental system e.g. $L \ge L_\mathrm{min}$ dictated by ion trap electrode distances, or the relationship between $R$ and $D$ defined by the total cavity mirror volume $V$.

Another important `controlled' parameter is the atomic branching ratio ($\alpha=\Gamma_1/\Gamma$, the ratio of the cavity arm transition decay rate to the total decay rate of the intermediate state $\ket{e}$) -- this value is set by the choice of atomic species and level scheme, and although not continuously tuneable, a range of values are possible. 

The second group of parameters are the `parasitics', all of which would ideally be zero. Here we assume that the system designer has limited control or knowledge of the actual value, but is able to establish an upper bound or tolerance on each. For our analysis of the ion-cavity system the parasitics we consider are the root-mean-squared roughness of mirror surfaces $\sigma$ and the misalignment $M$. There are naturally many parameters that are beyond the experimentalists' direct control such as noise on the magnetic field or laser intensity, however these have much wider impact than on the ion-cavity system alone, and lie beyond the scope of this paper.

The advantages of our two-stage optimisation scheme over na\"ive many-parameter optimisation schemes are readily apparent when we take into account the dimensionality of our parameter space. Given a certain set of parasitics, and constraints on the range of each controlled parameter, it is not unreasonable to perform a many-parameter optimisation of the photon extraction rate to identify the ideal cavity design. This family of optimisation schemes, provided a suitable algorithm is used, can provide both the optimal values for the controlled parameters and some estimate of the local shape of the cost function. While this provides a measure of how stable the system performance is against small perturbations, the broader landscape of the cost function remains opaque, preventing the identification of robust near-optimal operating regions. Alternatively, the full cost landscape may be exhaustively calculated, but this comes at considerable computational overhead, and ultimately the designer will likely end up making lower-dimensional cuts through the parameter space in a bid to understand the structure. In contrast with such global methods, the scheme we present in the following sections is both extremely simple and immediately yields the pertinent cuts through the higher dimensional parameter space. This additional information allows an experimentalist to select where to most efficiently expend effort to improve the parasitics with their system.

\subsection{$\mathcal{L}_\text{clip}$ and Optimisation of Cavity Geometry}\label{ssec:clipping}

\begin{figure}
    \includegraphics[width=\columnwidth]{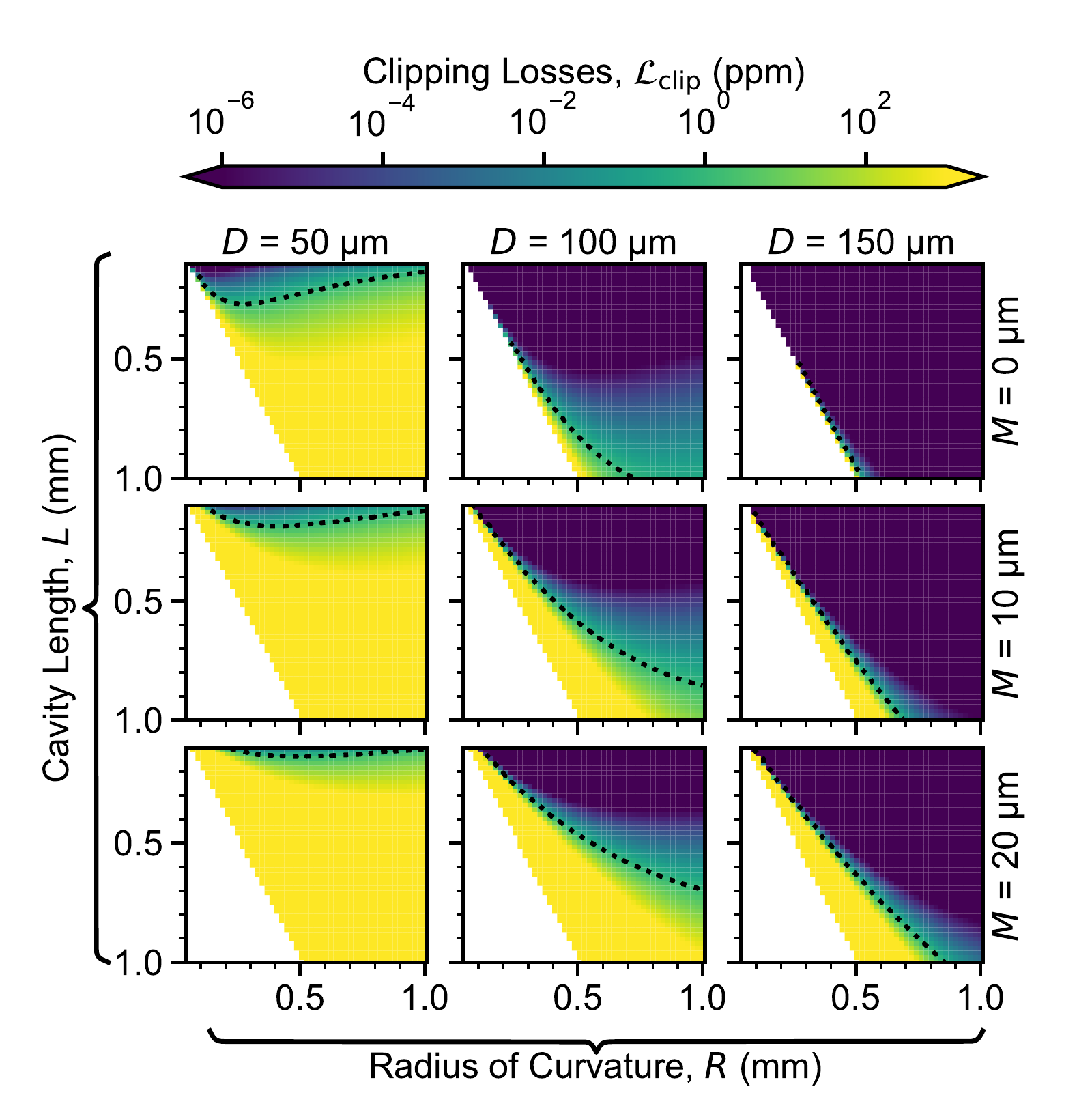}
    \caption{The dependence of round-trip clipping losses, due to cavity misalignment, on the cavity geometry. Each panel shows the value of $\mathcal{L}_\text{clip}$ as a function of the mirrors' radius of curvature and the cavity length for a given combination of mirror diameter ($D$; columns) and misalignment ($M$; rows). The white region indicates where the cavity does not support stable modes, or where the cavity cannot be assembled due to geometric constraints. Misalignment of the mirrors causes the mode to tilt at an angle $\phi$, with the effect more pronounced for cavities approaching the concentric condition ($R=L/2$). For small $\phi$, the mode remains well within the spherical region and losses are minimal, but as the mode approaches the edge of the mirror these rapidly increase. This leads to a clipping loss `threshold' which is particularly sharp for large and well-aligned mirrors. In all panels we show the 1 ppm clipping contour as a black dotted line. }
    \label{fig:clipping vs L R D M}
\end{figure}

\begin{figure}
    \centering
    \includegraphics[width=0.85\columnwidth]{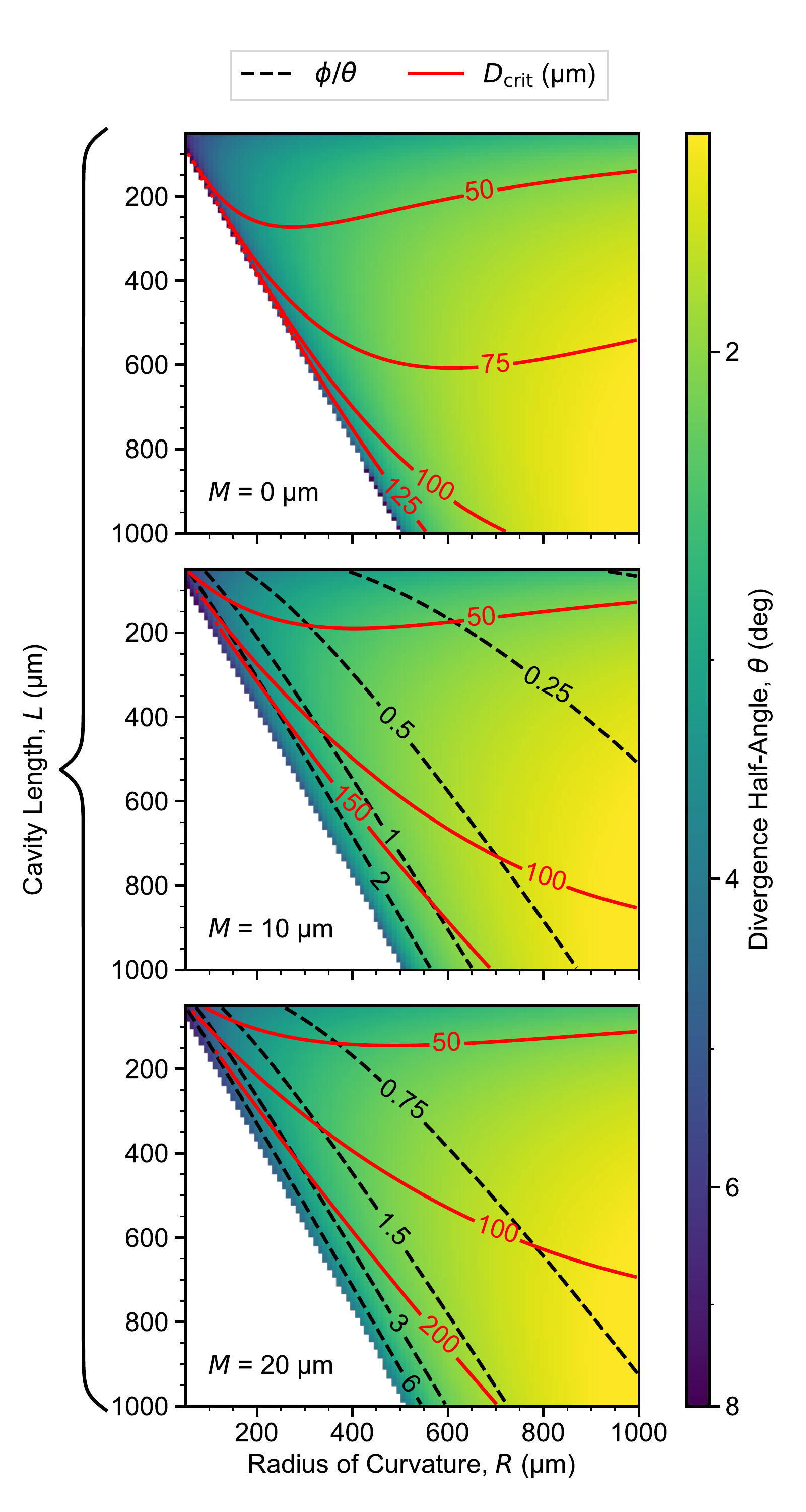}
    \caption{Maximising the geometric component of the cooperativity with finite-diameter mirrors. In each panel the colourmap shows the divergence angle $\theta$ as a function of the mirror radii of curvature and the nominal cavity length. $\theta$, and thus the geometric component of the cooperativity, increases strongly towards the concentric limit. The white region indicates where the cavity does not support stable modes. For three values of misalignment $M$, we show the minimum mirror diameter $D$ for which $\mathcal{L}_\mathrm{clip}$ is $1~\text{ppm}$ (red contours); and the ratio of tilt angle to divergence $\phi / \theta$ (black dashed contours). Note that the red contours are discontinuous where the cavities formed are geometrically limited. For a given diameter and misalignment, this allows us to identify the critical cavity geometry where clipping losses become significant, and to attribute the source of clipping loss to either mode tilt ($\phi \gg \theta$) or divergence ($\theta \gg \phi$).}
    \label{fig:Divergence angle optimization}
    \vspace{-0.5cm}
\end{figure}
We begin by optimising the geometric properties of the cavity ($L$ and $R$) to maximise the ion-cavity coupling rate while simultaneously minimising the loss due to clipping of the mode by mirrors of finite diameter $D$. Independent optimisations of these two metrics drive the cavity geometry in broadly opposing directions, so we must consider how to optimally trade performance in one sector against that in the other.

Our first step is to rewrite Eq~\ref{eq:C_in1} using Eqs~\ref{eq:lossin}, \ref{eq:theta}, and \ref{eq:g0}, to express this as a product of a purely atomic term, the cavity-arm branching ratio $\alpha$; a purely geometric term quantified by the divergence half-angle $\theta$; and the reciprocal of the intrinsic round-trip cavity loss $\mathcal{L}_\text{in}$:
\begin{equation}
    C_{\text{in}} =  6\alpha\theta^2\frac{1}{\mathcal{L}_\text{in}}.
\end{equation}
As discussed in Sec~\ref{sec:def}, the intrinsic cavity loss $\mathcal{L}_\text{in}$ is the sum of both clipping loss $\mathcal{L}_\text{clip}$ and mirror surface scattering loss $\mathcal{L}_\text{scat}$. In principle, to maximise $P_\text{ext}$ one needs to optimise the cavity geometry for both divergence $\theta$ and clipping loss $\mathcal{L}_\text{clip}$ simultaneously, which implicitly demands an involved procedure, accounting for $\mathcal{L}_\text{scat}$ and introducing mirror transmission as an additional dimension of optimisation. However, we will now show that the geometric element of the optimisation is well-approximated by the maximisation of a single parameter, $\theta$, within a boundary imposed by $D$ and $M$ which ensures $\mathcal{L}_\text{clip}$ remains small. The remaining optimisation of $\kappa$ via outcoupler transmission then has an analytic solution.

To calculate the clipping loss, we assume the small tilt due to misalignment causes a lateral shift, $\Delta x = \phi L'/2$, of the cavity mode at the mirror. As we are considering the perturbative regime of small clipping loss, and because we assume the cavity is operated far from transverse mode degeneracy, mode mixing effects can be ignored and the clipping is modelled simply as loss of amplitude~\cite{Clarke2018}. By extending the work of Hunger~\textit{et al.}~ \cite{Hunger2010} from the axial case: $\mathcal{L}_\text{clip}$ is approximated by considering the overlap of the shifted, unity-normalised $\text{TE}_\text{00}$ Gaussian mode $G'$ on the cavity mirror surface $S$:

\begin{equation}
\begin{split}
G' = G_0 e^{-\frac{(x-\Delta x)^2+y^2}{w^2}} &= G_0 \text{exp}\left[-\frac{(x-\phi \frac{L'}{2})^2+y^2}{(\frac{\lambda}{\pi \theta})^2 + (\frac{L'}{2}\theta)^2}\right],\\
\mathcal{L}_\text{clip}(L,R,M,D) &= 1-\int_{S(D)} |G|^2 \text{d}A.
\end{split}
\label{eq:clipping_loss}
\end{equation}
where $G_0$ is a normalisation factor, $w=w(\theta,\phi, L')$ is the cavity mode size at the cavity mirror surface, $\phi = \phi(M,L,R)$ and $\theta=\theta(M,L,R)$ are the tilt angle and the divergence half-angle.

Fig~\ref{fig:clipping vs L R D M} shows the dependence of $\mathcal{L}_\text{clip}$ on the cavity geometry parameters $L$ and $R$, for a range of mirror diameters $D$ and misalignments $M$. Naturally, clipping losses are higher for small diameter mirrors and large misalignments, but for many cavity geometries the tilted mode remains well-confined within the extent of the mirror and losses are negligible. However, as the cavity approaches the concentric configuration ($R=L/2$) the mode tilt and divergence rapidly increase, along with the resultant losses. This leads to a threshold effect which is particularly pronounced in the case of large diameter mirrors and small misalignment.

Noting this threshold, we can set a maximum tolerable loss $\mathcal{L}^\text{max}_\text{clip}$, defining a critical $R$-$L$ contour. On one side of the contour clipping loss will remain negligible for all $\{R, L\}$, but once crossed losses will rapidly increase to an unacceptable level. Qualitatively, this critical geometry describes how concentric one can safely make the cavity before mode tilt and divergence make the optical mode hit the edge of the mirror. Here we choose $\mathcal{L}^\text{max}_\text{clip} = 1~\mathrm{ppm}$, ensuring that clipping losses will be insignificant for even the highest-finesse cavities considered. We note that, whilst we have chosen an arbitrary threshold for our analysis, the exact value makes little difference to the overall properties of the cavity. This is because, for all but the smallest mirrors, the onset of clipping loss is so rapid as to make our ``critical contour'' nearly independent of $\mathcal{L}^\text{max}_\text{clip}$.

Within this boundary $\mathcal{L}_\text{clip}\ll\mathcal{L}_\text{scat}$, and so ${\mathcal{L}_\text{in}\approx\mathcal{L}_\text{scat}}$ which is independent of cavity geometry. The optimisation of clipping loss is thus reduced to a numerically calculated bound on the single parameter optimisation of $\theta$. Under this approximation the intrinsic cooperativity becomes:

\begin{equation}
    C_{\text{in}} \approx  \frac{6\alpha\theta^2}{\mathcal{L}_\text{scat}},
    \label{eq:approximated C_in}
\end{equation}
where the dependence of $\theta$ on cavity geometry is given by Eq~\ref{eq:thetaprime}.

Fig~\ref{fig:Divergence angle optimization} shows the dependence of $\theta$ on the cavity geometry in the presence of varying amounts of misalignment. The clipping loss boundaries for a range of mirror diameters are given as red contours, while the dashed black contours give the ratio of mode tilt to divergence angle, $\phi / \theta$, indicating which of these two effects dominates the clipping loss for a given configuration.

The optimal cavity geometry is therefore found by optimisation of $L$ and $R$ along the boundary associated with the available mirror diameter. While $\theta$ diverges at the concentric limit $L \rightarrow 2R$, this very divergence along with increased tendency for high $\phi$ means the clipping limit restricts how closely we can approach this condition. We note for completeness that $\theta$ also diverges for $L \rightarrow 0$, but there is always a practical limit on minimum cavity length, as discussed in Section~\ref{sec:intro}.

\subsection{$\mathcal{L}_\text{scat}$ and Optimisation of Mirror Transmission}\label{ssec:scattering}
\begin{figure}
    \centering
    \includegraphics[width=\columnwidth]{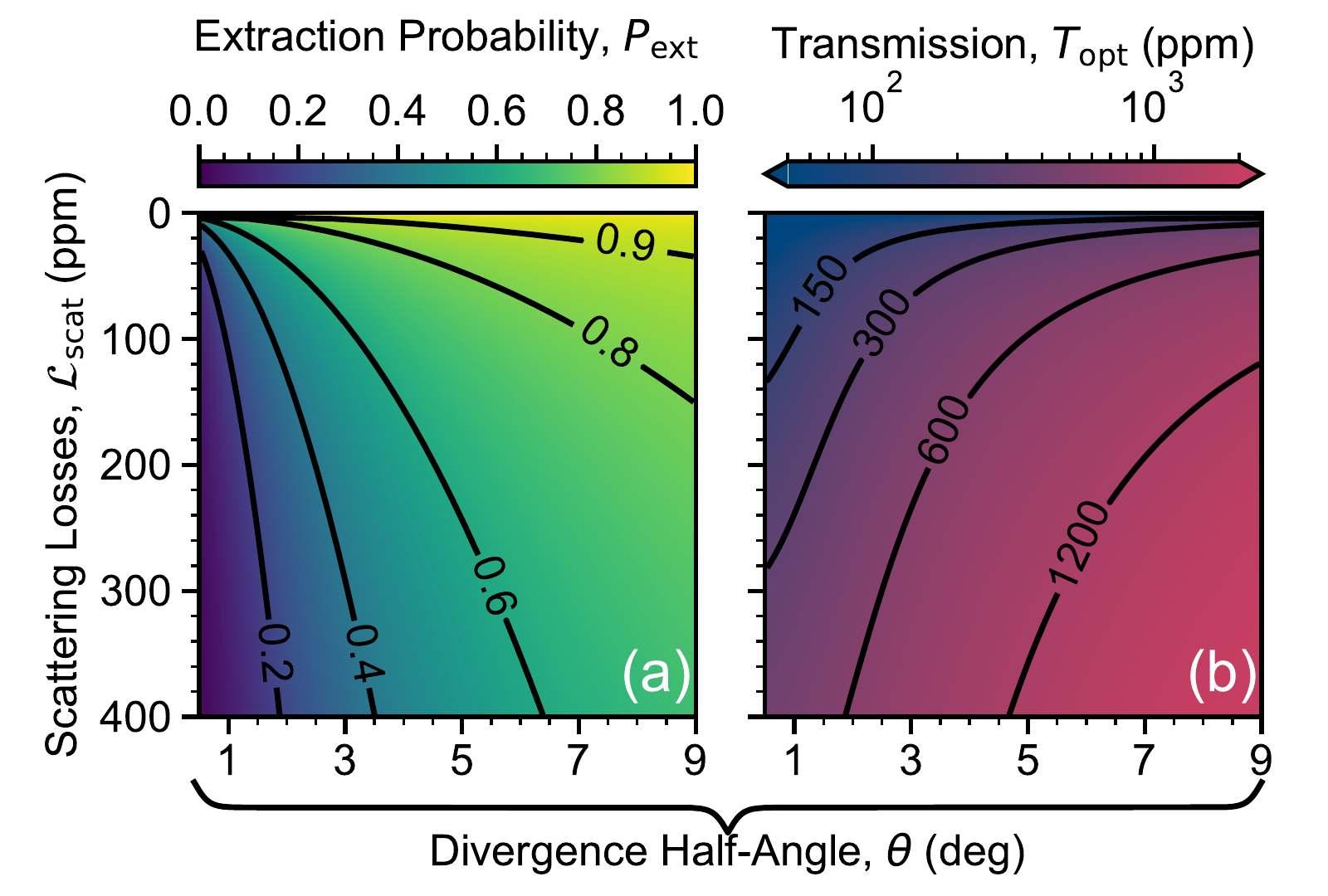}
    \caption{Performance optimisation vs mode divergence and round-trip scattering loss. (a) Optimised photon extraction probability, $P_\text{ext}$. The divergence angle is set by the geometry, as shown in Fig~\ref{fig:Divergence angle optimization}, whilst the scattering losses $\mathcal{L}_\text{scat}$ are fixed by the mirror manufacturing process. We have assumed an atomic transition branching ratio of $\alpha=1/20$. (b) The corresponding optimal output coupler mirror transmission, $T_\mathrm{opt}$, needed to achieve the value of $P_\text{ext}$ in (a). The contours in each panel follow the colourmaps.}
    \label{fig:parameter vs theta and scatter}
\end{figure}

In this Section we consider the impact of the optical properties of the cavity mirror itself, specifically the scattering losses due to surface roughness $\mathcal{L}_\text{scat}$ and the transmission coefficient of the outcoupler, $T$. The geometric optimization in Section~\ref{ssec:clipping} gives the maximum possible $\theta$ and thus, through Eq~\ref{eq:approximated C_in}, also maximises $C_{\text{in}}$. Maximising $P_\text{ext}$ now depends upon choosing a suitable value of $T$. We desire that photons generated within the cavity are transmitted through the outcoupler rather than scattered, achieved in the limit $T\gg\mathcal{L}_\text{in}$. However, applying this limit means $C\ll C_\text{in}$, which will reduce $P_\text{gen}$. Achieving optimal performance requires us to balance these two considerations. With $\mathcal{L}_\text{clip}\approx0$ after geometric optimisation, the transmission optimisation now depends only on the scattering loss. Assuming that there is transmission only through the outcoupler mirror, the optimal value for this parameter can be determined analytically (following \cite{Goto2019}):
\begin{equation}
    T_\text{opt} = \mathcal{L}_\text{in} \sqrt{1+2C_\text{in}}\approx \mathcal{L}_\text{scat} \sqrt{1+2C_\text{in}}.
    \label{eq:transmission optimisation}
\end{equation}
We apply this approximation and Eq~\ref{eq:probability} to optimise $P_\text{ext}$ for a range of values of both $\theta$ and $\mathcal{L}_\text{scat}$, the results of which are shown in Fig~\ref{fig:parameter vs theta and scatter}. As can be seen from the figure, for larger $\theta$ it is possible to achieve higher extraction efficiencies for a given $\mathcal{L}_\text{scat}$ by increasing mirror transmission. However, it is important to note that when scattering losses are sufficiently low, the high cooperativity saturates $P_\text{ext}$, making performance in this regime only weakly dependent on cavity geometry.

\subsection{Combined Optimisation of $P_\text{ext}$}\label{ssec:combined}

In this Section we combine the results of Sections~\ref{ssec:clipping} and \ref{ssec:scattering} to establish an optimal approach to cavity design in the presence of known engineering tolerances.

\begin{figure}
    \centering
    \includegraphics[width=0.9\columnwidth]{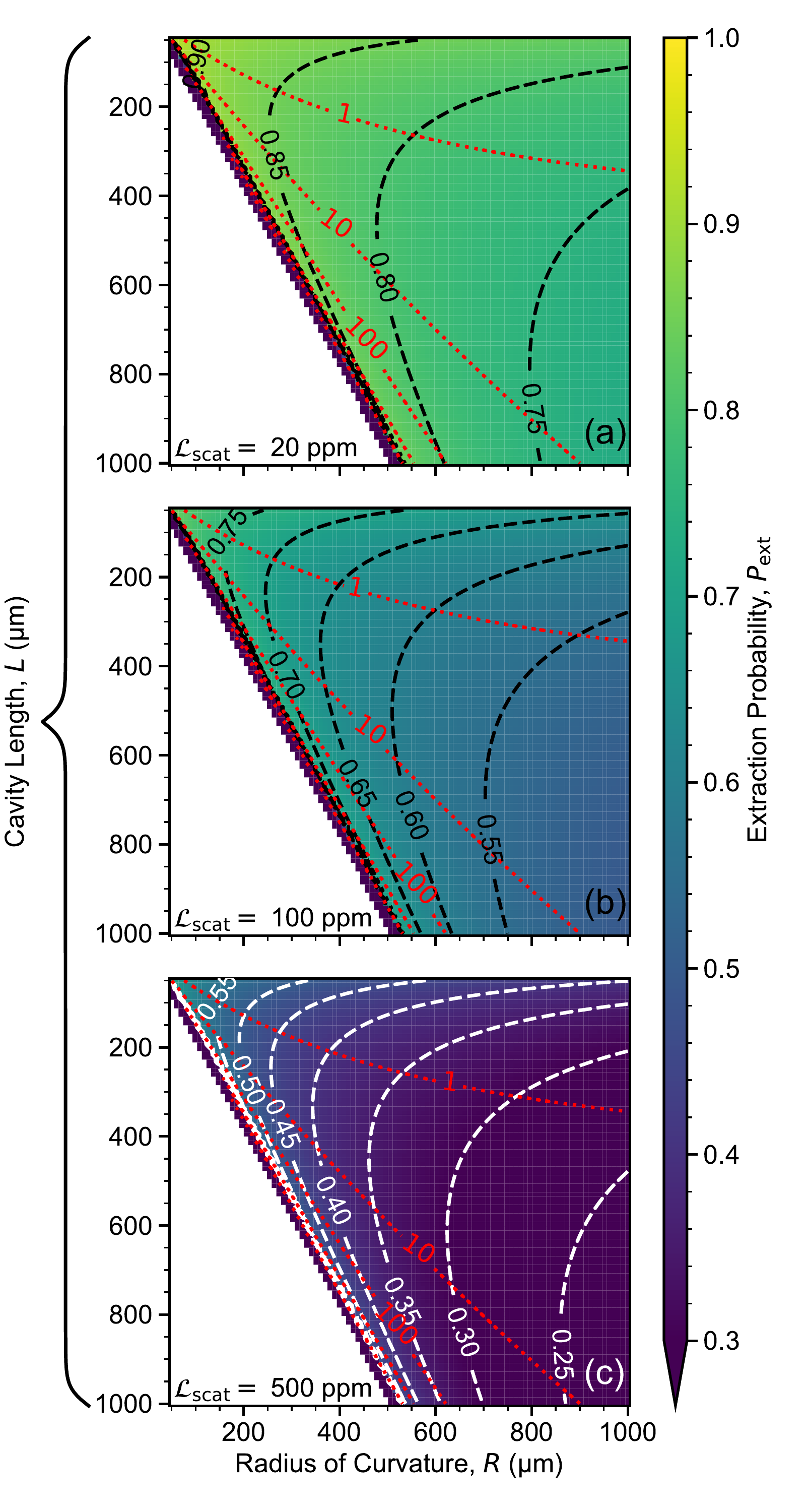}
    \caption{ The parameter space searched for the optimal ion-cavity system. The colour mapping shows performance as a function of cavity length $L$ and mirror radius of curvature $R$, for fixed misalignment $M=\SI{5}{\micro\metre}$ and atomic branching ratio $\alpha=1/20$. Plots (a)-(c) correspond to the cases with round-trip scattering loss $\mathcal{L}_\text{scat}=\{20, 100, 500\}\,\text{ppm}$ respectively. The dashed black (white) contours show the optimised photon extraction probability, $P_\text{ext}$. Dotted red contours in (a) indicate the critical value of the mirror volume, $V$, in pL ($1~\mathrm{pL} = 10^{-15}~\mathrm{m}^3$) while the white region indicates where the cavity mode is unstable. Note that the stability region and the iso-volume contours are consistent across all three configurations. }
    \label{fig:parameter vs geometry}
\end{figure}

Examples of typical geometric optimization landscapes are given in Fig~\ref{fig:parameter vs geometry} (a)-(c) for three different values of scattering loss and a constant misalignment of $M=\SI{5}{\micro\metre}$. In each panel, we show the optimum value of $P_\text{ext}$ for a given $R$ and $L$ as a heatmap, with accompanying dashed black contours, while the dashed red contours in (a) show contours of fixed mirror volume, $V$, for which clipping losses, as defined in Sec~\ref{ssec:clipping}, are fixed to 1~ppm. These plots resemble the divergence angle heatmap of Fig~\ref{fig:Divergence angle optimization}(b), but now show how these translate to overall $P_\text{ext}$ performance when combined with a range of $\mathcal{L}_\text{scat}$. Note that while the qualitative behaviour is similar in each case, in the presence of higher scattering losses the efficiency is both lower and more strongly dependent on the cavity geometry. We note that this optimum value of $P_\mathrm{ext}$ depends heavily on the value of the atomic branching ratio $\alpha$ through the cooperativity. As shown in Eq.\ref{eq:approximated C_in} $C_\mathrm{in}\propto\alpha$ and $\alpha$ is independent of the cavity geometry so this is a simple re-scaling of the maximum achievable photon extraction probability. Hereafter, for the purposes of numerical evaluation, we fix the value of the branching ratio to $\alpha=1/20$, which appropriately matches schemes in alkali-like atoms and singly-ionised alkaline-earth-like atoms where $\ket{0}$ is a stable $S$ state, $\ket{e}$ is an short-lived $P$ state and $\ket{1}$ is a meta-stable $D$ state.

We can see from Fig~\ref{fig:parameter vs geometry} that $P_{\text{ext}}$ is maximized at different geometric positions depending on the clipping loss boundary, set by mirror volume $V$. By following the iso-volume contours, in the direction of decreasing $L$ the extraction probability increases. The maximum $P_{\text{ext}}$ thus lies at the top left corner of the region bounded by the clipping loss boundary and the minimum cavity length. Therefore, when there is no further constraint on $R$, the optimal cavity length is \emph{always} the minimum length compatible with the experiment in question.

In some situations the assumption of an unconstrained optimisation of $R$ will not hold. This might occur if using pre-existing mirrors with fixed values of both $R$ and $D$, or be a result of the challenges of making mirrors with low radii of curvature or with large diameters. In a typical case where $R\simeq L_\text{min}$ and both $R$ and $D$ are fixed, the optimal length is instead the longest permissible without exceeding the clipping loss threshold~\footnote{In the rarer case of $R \gg L_\text{min}$ combined with small $D$ and/or large $M$, minimising $L$ can again be the best strategy, however overall performance in this regime tends to be poor and we do not consider it further.}. However, as we will later show in Sec~\ref{ssec:robustness}, the resulting increase in $P_\text{ext}$ is fairly limited, and for robust operation it may still be preferable to use ${L=L_\mathrm{min}\approx R}$ in this scenario.

\section{Optimisation subject to constraints}\label{sec:perform}
In Sec~\ref{sec:optimise} we determined a strategy for optimising photon extraction probability in the presence of limitations, such as scattering loss and lateral misalignment of the cavity mirrors. In this Section we will discuss the expected performance of systems optimised following this strategy. In Sec~\ref{ssec:parameterlimits} we consider such an optimisation, subject to limitation on the cavity length and volume of the mirror. In Sec~\ref{ssec:robustness} we consider the impacts of inaccurate estimation of the manufacturing tolerances and deviation from the optimal design parameters to quantify the robustness of the optimised system.

\subsection{Performance vs. Parameter Constraints}\label{ssec:parameterlimits}

 \begin{figure*}
     \centering
     \includegraphics[width=0.95\textwidth]{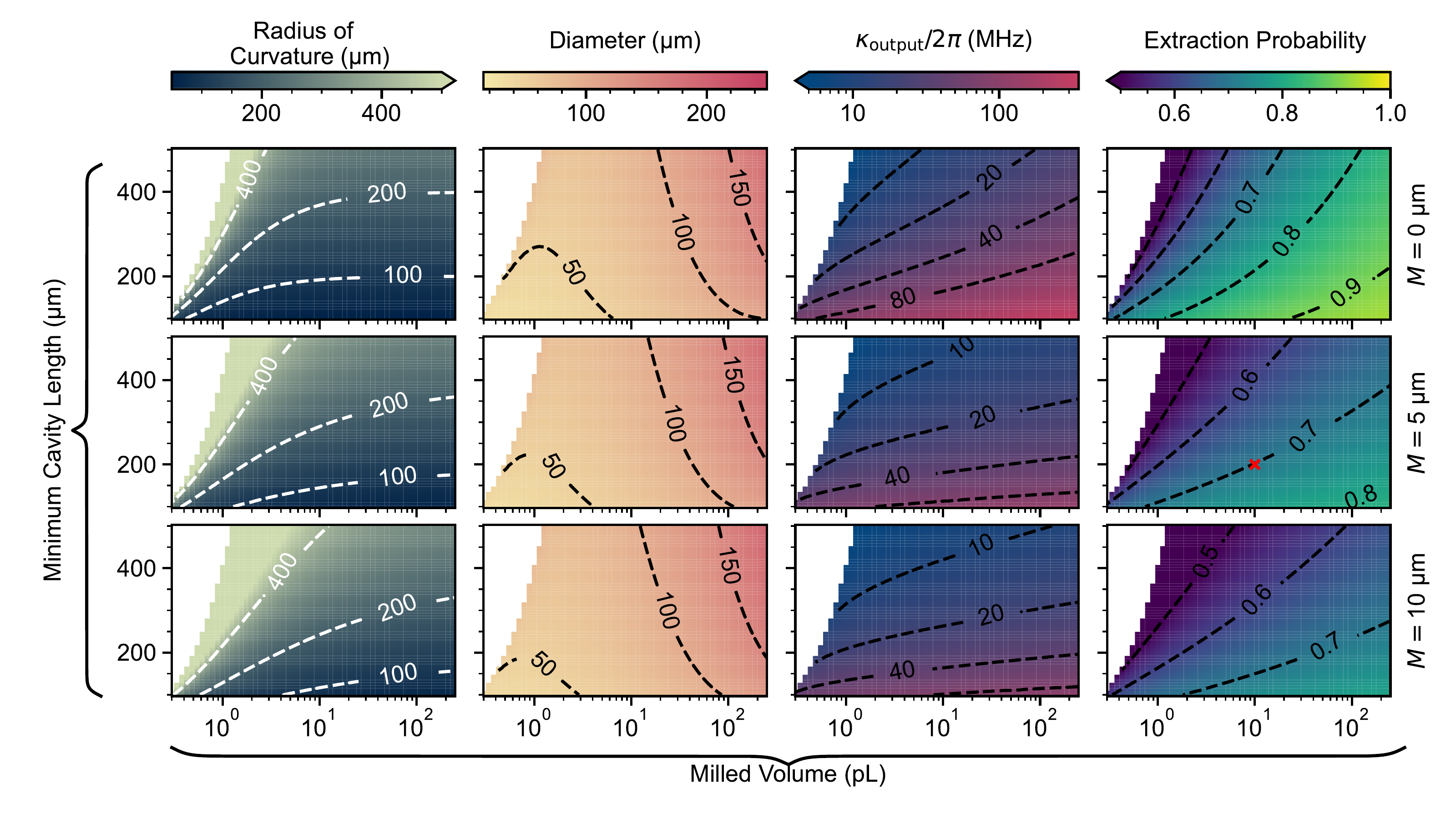}
     \caption{Optimum cavity performance for fixed milling volume and minimum lengths. For all configurations shown the value of scattering loss is fixed to 100 ppm. Regions where the cavity optimiser fails to find a suitable optimum, subject to these constraints, are shown in white. The red cross ($\times$) shows the geometry considered in Fig.~\ref{fig:robustness}.}
    \label{fig:optimum}
 \end{figure*}

Previously, we have placed no explicit limitations on the geometry of the cavity system, other than requiring that the mirrors be physical and the resonator be stable. In practice however, this is not achievable. In any realistic experimental scenario there will be additional constraints applied beyond simple geometrical limits. In this subsection we explore how the optimal cavity is impacted by experimentally motivated limits on mirror volume and minimum cavity length.

In the absence of any constraints, the optimal geometry will tend towards $(L, R)\rightarrow0$, but as discussed in Sec~\ref{sec:intro}, such a geometry is neither practical nor physical. Due to the presence of other experimental features within the apparatus, e.g. ion trap electrodes or focused laser beams, there is usually a minimum length that one can realistically achieve before impacting the overall functionality of the experiment. In Fig.~\ref{fig:parameter vs geometry} we show the parameter space over which one would optimise a cavity with no external constraints on $L$ or $R$. Additionally, as dotted red contours, we show how the parameter space is further constrained when we consider the additional external constraints of fixed mirror volume and $\mathcal{L}_\mathrm{clip}\leqslant 1~\mathrm{ppm}$. Combining this limitation with a minimum length constraint provides a unique maximum. To avoid biasing our results we allow the optimiser to freely vary $L$. We find that, within the region that we explore, the optimal mirrors have a saggita that is significantly less than the minimum length of the cavity. Furthermore, the optimal length is always the minimum imposed, with the radius of curvature tending towards concentricity. Ultimately the optimal cavities are often far from the concentric limit, due to the fixed mirror volume and the necessity of avoiding excessive clipping loss. As the available values of $R$ and $D$ are inter-related via $V$, movement towards lower clipping losses through an increase in $D$ driving a requisite increase in $R$, for reducing the divergence angle and moving the cavity away from concentricity.

In each of panels (a), (b) and (c) of Fig.~\ref{fig:parameter vs geometry} we show the same geometries but for progessively higher scattering losses (20~ppm, 100~ppm and 500~ppm) respectively. As is to be expected there is no change in the general properties of the cavity, with regards to the clipping loss boundaries and the stability criterion. Crucially, for a given volume, the peak extraction probability occurs for the same geometry no matter the value of the scattering losses. However, as scattering losses there is a significant decrease in the value of this maximal extraction probability.

Having identified the four parameters that define the optimisation landscape, $V$, $L_\mathrm{min}$, $M$ and $\mathcal{L}_\mathrm{scat}$, we now explore the full four-dimensional parameter space. However, following from our preceding analysis, we have established that $\mathcal{L}_\mathrm{scat}$ simply reduces the maximal value of $P_\mathrm{ext}$ and the optimal value of transmission. As such we are able to restrict ourselves to presenting results for a single value of $\mathcal{L}_\mathrm{scat}$ without changing the optimal cavity geometry presented. We have also already established, and shown in Fig.~\ref{fig:Divergence angle optimization}, that increasing misalignment corresponds to introducing non-zero $\phi$ and modification of the position of clipping loss contours. Naturally this does lead to changes in the presented optimal geometry as well as the maximal value of $P_\mathrm{ext}$. Numerically however, we find that is a slow variation across a small range of parameters. As such we can consider variation across a limited range of discrete values of $M$. Across the two remaining parameters, $L_\mathrm{min}$ and $V$, we expect a more complex trade-off for optimal $P_\mathrm{ext}$ as such we present these as continuous axes.

In Fig.~\ref{fig:optimum} we show the parameters related to the optimum cavity for a given pair of $L_\mathrm{min}$, the minimum length, and $V$, the milling volume. To determine the optimum we use a Nelder-Mead algorithm with a random pair of initial parameters $L, R$ chosen within the stability region. To prevent false local optima from being presented we repeat this process with multiple initial conditions until a pair of optima are returned with less than 0.1\% variation in the objective function. To aid convergence we limit the maximum values of $L$ and $R$ to 3~mm. 

Considering first the panels showing $R$ and $D$: we see that for all but the lowest milling volumes and longest lengths it is advantageous to maintain $R \ll 500~\mathrm{\upmu m}$. This increases the mode divergence and gives a smaller mode waist to increase coupling - however as the cavity length is kept short the mode size at the mirror is small enough to prevent clipping losses. As maximum volume increases it is more optimal to increase $D$ than it is to decrease $R$, this is because the cavity is designed to operate close to concentricity, changes in $R$ then lead to rapid onset of clipping losses. 

A key consideration for using cavities in quantum networking is the rate at which a photon can be effectively extracted from the cavity, this is governed chiefly by $\kappa_\mathrm{output}$ which is related to transmission, $T$ by
\begin{equation}
    \kappa_\mathrm{output} = \frac{T c}{4 L}.
\end{equation}
In Appendix~\ref{app:Adiabatic} we show that for, reasonable vSTIRAP pulse parameters, the adiabatic limit can be achieved with pulses of duration $\sim 10/\kappa$. The lowest transmissions we find are for cavities with long minimum lengths, small mirror volumes and high misalignment; these yield values of $\kappa_\mathrm{output}/2 \pi \approx 10 ~\mathrm{MHz}$ or extraction times of order $1~\mathrm{\upmu s}$. As discussed in Sec.~\ref{ssec:rate}, for the majority of experiments attaining an attempt time on the order of $1~\upmu\mathrm{s}$ is challenging due to necessary state preparation or cooling steps, as such we do not consider these low $\kappa$ solutions to be inherently limiting.

\subsection{Robustness of Performance}\label{ssec:robustness}

\begin{figure*}
    \centering
     \includegraphics[width=0.9\textwidth]{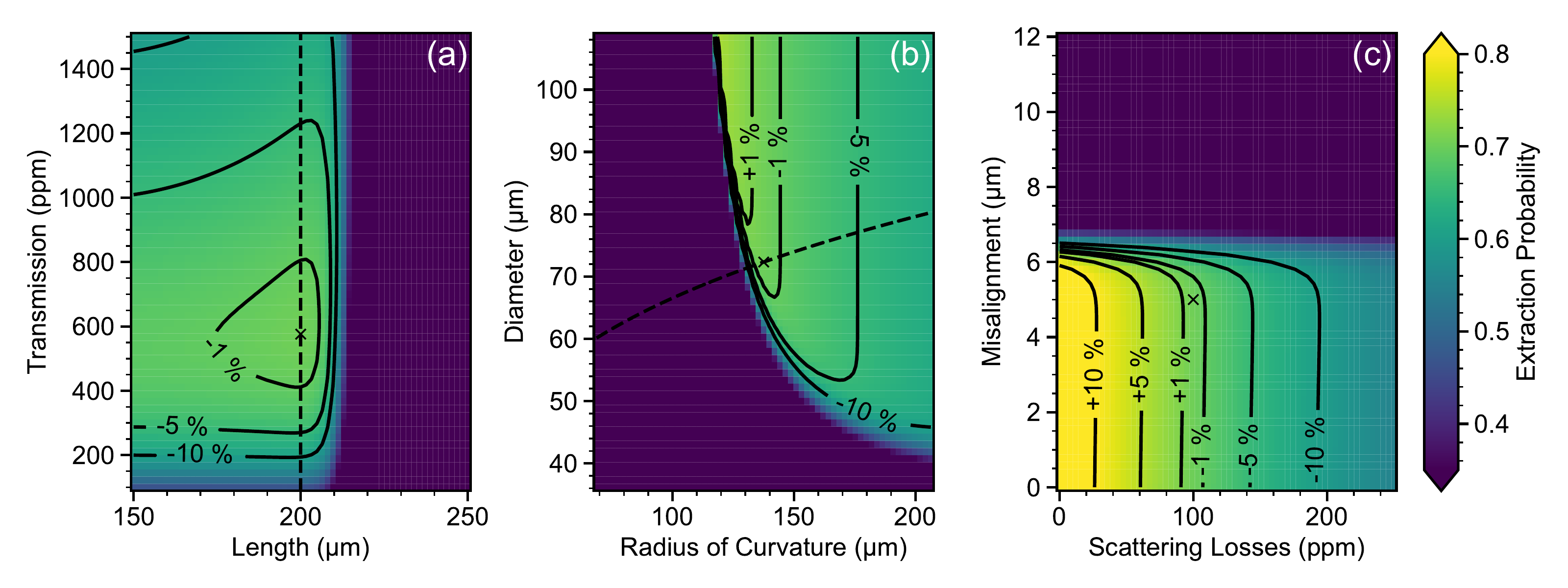}
     \caption{Performance of an optimised ion-cavity system subject to manufacturing defects. In all three panels the extraction probability is shown by the colour map, the cross ($\times$) represents the optimum set of parameters and the solid black contours show the $\pm 1\%, 5\%$ and $10\%$ levels in extraction probability. The cavity mirrors have a fixed design volume of $10~\mathrm{pL}$, $\mathcal{L}_\mathrm{scat} = 100~\mathrm{ppm}$ and $M = 5~\mathrm{\upmu m}$. (a) The cavity extraction probability as a function of cavity length and transmission, keeping $R$ and $D$ constant. The dashed vertical line indicates the fixed value of the minimum length used to optimise the mirror geometry. We see that the maxima is not strongly constrained by $T$, however for a $\sim5\%$ increase in $L$ the extraction probability has dropped away; the converse is not true with a 1\% error on $P_\mathrm{ext}$ for approximately a $10\%$ reduction in $L$. (b) Changes in $P_\mathrm{ext}$ due to mirror fabrication, with the volume constraint lifted. The dashed black line is the iso-volume contour for the design volume. Increasing $R$ degrades performance rapidly, whilst changes that decrease $R$ lead to unstable operation. Increasing $D$ has only a small impact on performance, while decreasing $D$ also leads the cavity towards high clipping losses and poor extraction probabilities.  (c) Change in extraction probability due to poor estimation of $M$ and $\mathcal{L}_\mathrm{scat}$. Underestimates of $M$ result in unstable cavity configurations, whilst underestimates of $\mathcal{L}_\mathrm{scat}$ simply degrade performance. Overestimates of $\mathcal{L}_\mathrm{scat}$ improve performance, however there is no such change for overestimates of $M$. }
    \label{fig:robustness}
 \end{figure*}

While the optimum value for both $M$ and $\mathcal{L}_\text{scat}$ is zero, we have optimised our system on the basis of an assumed achievable value for these parameters. These tolerances are only estimated before manufacture, and in the case of $M$ the true value will not be known until the cavity is operational. It is therefore crucial to consider the impact of incorrectly estimating these tolerances and to ensure that small errors of this sort will not substantially degrade the performance. In this Section we first consider a system in which cavity length, outcoupler transmission and the mirror geometry are optimised, as shown previously, to maximise $P_\text{ext}$ for a fixed values of $M$, $\mathcal{L}_\text{scat}$, $V$ and $L_\mathrm{min}$. We then consider how the performance of this optimised system is impacted if the estimates of $M$ and $\mathcal{L}_\text{scat}$ used during the optimisation are incorrect and how manufacturing tolerances on the mirror assembly, through $L$, $T$, $R$ and $D$, impact performance.

In Fig.~\ref{fig:robustness}(a), we first consider how variations in length and transmission affect the cavity performance. Unsurprisingly, if the cavity is made longer than the design length, which in this case is also the minimum allowed, then the cavity enters a region of poor performance operation, with clipping losses rapidly reaching greater than the 1~ppm threshold we have previously imposed. In the converse case; in which the cavity is made shorter than designed (putting aside that our limitation was due to performance of other aspects of the system) we observe a more gradual drop off in the performance of the cavity. However, even in the more sensitive direction (where $L > L_\mathrm{opt}$) the performance remains at 99\% of the optimal value for a 3\% change in length. Considering now the effect of transmission, at the optimal length the 1\% degredation contour spans nearly 400 ppm change in transmission, requiring only a 30\% tolerance, far in excess of typical tolerances on high-quality dielectric coatings - which are typically specified to a few ppm.

The other important cavity manufacture parameters are those related to the mirror geometry: $R$ and $D$. In Fig.~\ref{fig:robustness}(b) we show the parameter space spanned by these two parameters, noting that we have lifted our constraint that the milled volume is fixed, instead indicating this limitation with a contour of constant $V$. The dependence of $P_\mathrm{ext}$ on diameter is minimal, demonstrating the rapid onset of clipping losses, as shown first in Fig.~\ref{fig:clipping vs L R D M}. Additionally we see a general increase in $P_\mathrm{ext}$ towards smaller values of $R$, again bounded by clipping losses. However because of our constraint on milled volume, the optimum is found along a contour that is near perpendicular to that for clipping loss. This implies that the trade-off for reduced sensitivity to clipping is a loss of $P_\mathrm{ext}$. As such, when subject to the constraint on maximum volume, the opportunity cost of conservative mirror design i.e. larger diameters, is significant in available extraction probability. For instance: a 5 ~$\mathrm{\upmu m}$ increase in diameter from optimal causes a 5\% drop in extraction probability due to the large increase in $R$ needed to maintain the constraint on design volume.

Finally, we consider the possibility that the estimates made of the values of $M$ and $\mathcal{L}_\mathrm{scat}$ are incorrect. This is not an unrealistic scenario, as both parameters need to be estimated \textit{a priori}. As is to be expected, when both parameters are initially underestimates, that is: the true value of $M$ and $\mathcal{L}_\mathrm{scat}$ are greater than used for the optimisation, the cavity performance is degraded. With increased $\mathcal{L}_\mathrm{scat}$ the cavity remains stable but with reduced extraction probability, however for $M$ the cavity rapidly loses performance as it approaches a region of large clipping losses. In the converse case, where the two parameters are overestimates, the cavity continues to function at, or beyond, its design level. With lower $\mathcal{L}_\mathrm{scat}$ the extraction probability increases, there are no performance gains however for overestimating $M$ as clipping losses have already been made negligible, further reductions provide no gains. From this analysis we conclude that making accurate estimates of $M$ is critical when building static monolithic cavities, and so to ensure stable performance the value used for optimisation should be the 2-$\sigma$ upper bound or greater however state-of-the-art performance is achievable within expected experimental tolerances.

\section{Summary and Outlook}\label{sec:outlook}

In this manuscript we have considered how the design of `real-world' ion-cavity systems can be tailored to optimise their performance for quantum networking applications, in particular the generation of single photons from ions for measurement-based remote entanglement schemes. We began by defining several figures of merit for the performance of such schemes. We showed that through suitable choice of ion species, photon generation scheme and driving pulse, it is possible to ensure that the impact of the cavity design on entanglement fidelity and attempt rate can be made negligible. This allowed us to identify a single figure of merit for our optimisation, namely the probability of photon extraction from the cavity per attempt cycle, $P_\text{ext}$.

To optimise $P_\text{ext}$, we showed that in this context the cooperativity can be well approximated as a product of three independent components, depending respectively on cavity geometry, the electronic structure of the ion, and scattering from the mirror surface. The separability of the problem allowed us to perform the optimisation process in a serial manner, first numerically optimising the geometry (for a given set of limits) before identifying the optimal mirror parameters analytically. The significant reduction in the dimensionality of this optimisation problem versus the more general case helps provide intuitive demonstrations of the trade-offs inherent in such ion-cavity system designs, and highlights regions likely to yield robust performance.

We explored these trade-offs subject to realistic constraints, coming from minimum cavity length and maximum available milling volume. By considering radii of curvature and diameter of mirrors to be freely exchangeable for a fixed volume, we find that the optimum microcavity is always the shortest permitted. The cavity mirror radius of curvature then tends towards the concentric limit ($R\rightarrow L/2$) with the diameter increased to prevent clipping at the 1~ppm level.

Providing the maximum transverse misalignment of the cavity mirrors is well-characterised and accounted for, we have shown that very high extraction probabilities can be achieved. The relaxation of alignment requirements makes possible a range of much simpler techniques for the fabrication of microcavities, an important step on the route to scalable quantum networks.

By studying the robustness of these optimal configurations with respect to errors in the machining and uncontrolled parameters we find that, should one's goal be to maximise extraction probability: efforts should always be expended in reducing scattering losses but residual misalignment can be accounted for at the design level, provided it can be predicted. Additionally we find that there is little reward for conservative mirror design with respect to clipping losses, as performance rapidly degrades as the radius of curvature moves from optimum. More stable operation can be achieved by increasing the diameter of the mirror, but at a significant opportunity cost.

Our conclusions regarding `optimal' cavity design do not, broadly speaking, contradict contemporary trends, but the significant range of geometries and operating regimes seen in groups around the world indicates that the field is some way from consensus on the topic. We intend our work to help other researchers to approach this multifaceted problem systematically, adapting our approach to the parameter regime defined by their experimental constraints. We have made our simulation code available on \footnote{The code used in this work is available at DOI:{10.5281/zenodo.7020047}} and encourage others to use and modify it towards the design of their own systems.

\section{acknowledgements}
The authors would like to thank Peter Horak for stimulating discussions. This work was funded by the UK Engineering and Physical Sciences Research Council Hub in Quantum Computing and Simulation (EP/T001062/1) and the European Union Quantum Technology Flagship Project AQTION (No. 820495).

The data and analysis presented is available \underline{from [DOI to be added at acceptance]/ at reasonable request}.

\bibliographystyle{apsrev4-2}
\bibliography{reference}
\appendix
\section{Temporal Mixing}\label{app:Mixing}
For many quantum information applications, it is important that the photons extracted from the cavity are indistinguishable from each other. Where this is the case, care must be taken in the design of the vSTIRAP-type schemes we consider, where spontaneous emission to the initial state during the photon production process can significantly enhance distinguishability.

The mechanism for this process is described in detail in~\cite{Walker2020, gao2020}, but can be summarised as follows. During the application of the Raman driving pulse, there is a finite chance that the ion will spontaneously emit a photon into free space and decay to the initial state, $\ket{0}$, interrupting the population transfer to the target state $\ket{1}$. Once in this initial state, the vSTIRAP transfer begins again, now driven by the remaining portion of the driving pulse (which may result in cavity emission with or without one or more further spontaneous emission events). Note that if spontaneous emission does not occur directly to $\ket{0}$ or to another state which itself rapidly decays to $\ket{0}$, then the transfer process terminates; this has an impact on average production rate but not on indistinguisability and is generally far more tolerable.

In each photon production attempt resulting in cavity emission, the emitted photon has a waveform determined by the time of the last spontaneous emission event (relative to the start of the driving pulse). On average the emission of the cavity is a mixture of these photons, weighted by the probability of observing the corresponding final spontaneous emission time. Photons produced after spontaneous emission exhibit distorted and retarded photon wavepackets, with the effects more severe the later the final spontaneous emission event occurs. This means that two photons from the same system may have highly distinguishable temporal wavepackets.

Combating this source of error requires that the last spontaneous emission times do not vary, or that spontaneous emission followed by cavity emission happens very rarely. There are three main ways to achieve this
\begin{enumerate}
    \item Increasing the cavity cooperativity: Increasing $C$ will increase the ratio of cavity emission to spontaneous emission, increasing the likelihood that the cavity photon is emitted without a spontaneous emission event during the production process and thus decreasing distinguishability. Increasing the cooperativity is, however, challenging in the presence of misalignment and finite mirror size (as discussed in the main body of the paper from the perspective of increasing total extraction), and therefore increasing the cooperativity to levels where this error is insignificant may not be technically feasible. Even where possible, this adds additional complication to the cavity optimisation process, involving trading off indistinguisability and production rate, a balancing act that depends greatly on the specific system and application.
    \item Fast excitation: If the excitation pulse is fast compared to the spontaneous lifetime of the excited state and cavity decay time (for instance by utilising a picosecond pulsed laser), spontaneous emission and subsequent cavity emission can only occur during this very short excitation window and all photons emitted from the cavity effectively have the same starting time. Note that in the case of fast excitation, production of the photon no longer proceeds via a vSTIRAP process, and the corresponding $P_\text{gen}$ is lower and obeys a distinct expression~\cite{gao2020}; if such a scheme is used, the optimisation steps in this paper would need to be significantly modified.
    \item Using a system with a favourable branching ratio: If the levels for the scheme are chosen such that the branching ratio to the initial state is much lower than that to the final state, i.e. $\Gamma_{0}\ll\Gamma_{1}$, the temporal distinguishability introduced by this process can be rendered negligible at even modest cooperativities. In this case, there is no need to drive the system very quickly to avoid the degradation of indistinguishability, and an adiabatic drive can be used without problems.
\end{enumerate}

In our own experiments, we plan to use a $\Lambda$-system that satisfies the third of these criteria, and thus need not consider issues of indistinguishability further during the optimisation of our design. For the purposes of the work presented in this manuscript we assume that this criterion has been met; we believe that efficient and scalable systems will be built on intrinsically robust production schemes and that this matter should be addressed independently of the optimisation of extraction efficiency.

\section{Adiabatic Driving}\label{app:Adiabatic}

\begin{figure}
    \centering
    \includegraphics[width=\columnwidth]{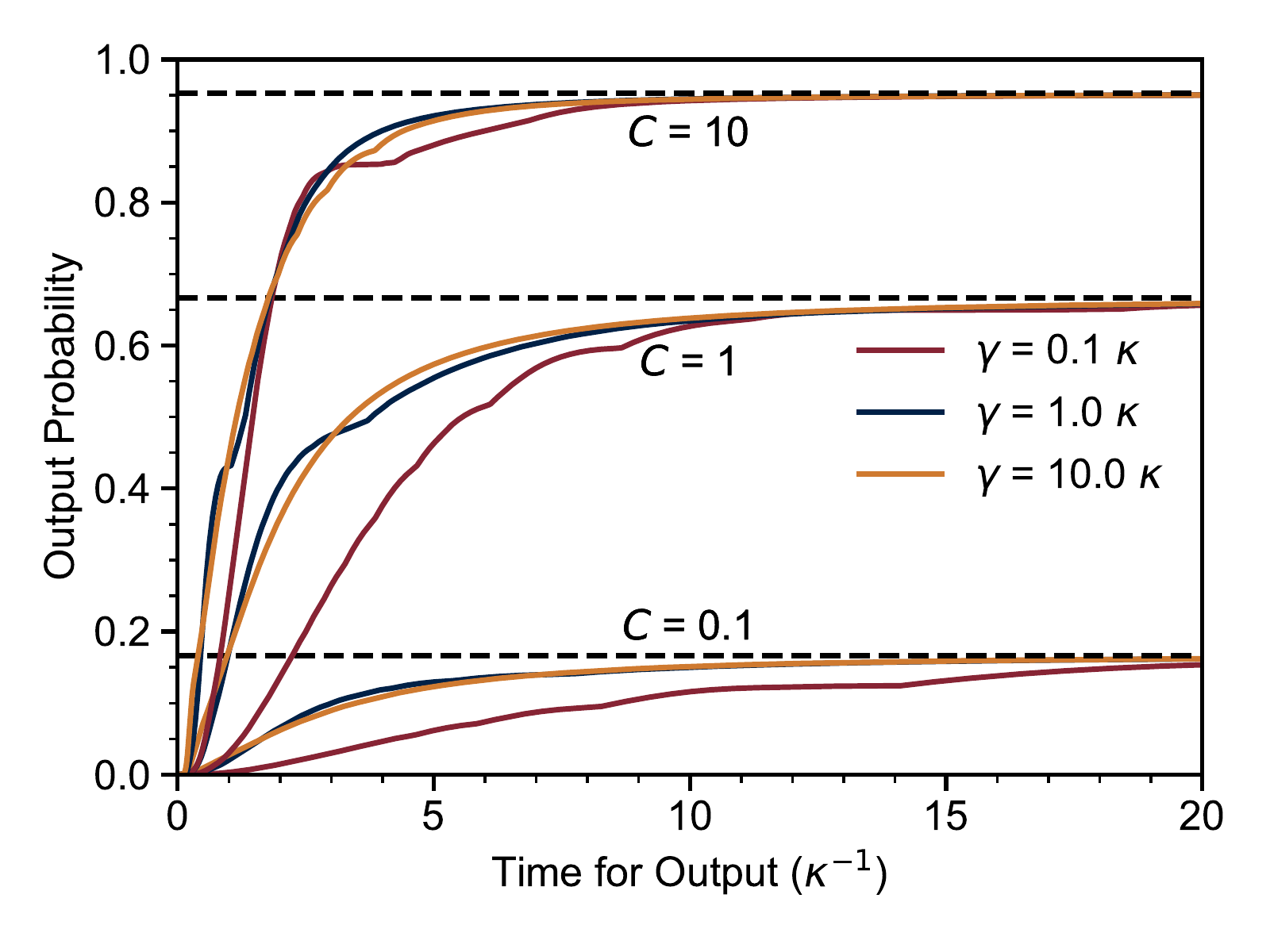}
    \caption{Plot of finite time photon outputs for a range of ion-cavity systems. The evolution of each system is simulated with a selection of pulses whose amplitude follows a sine-squared shape and is always on resonance with the $\ket{0}\rightarrow\ket{e}$ transition. At each time point, the largest value that the output probability takes over the range of simulated pulses is recorded as an example of a lower bound on the achievable output. The systems simulated form three groups with cooperativities of: $C=0.1$, $1.0$ and $10.0$, where the adiabatic limit $P_\text{gen}=2C/(2C+1)$, valid for slow driving/infinite output time window, is indicated with a dashed black line. Within each set $\gamma$ takes three different values to compare across a range of systems; note that for system configurations close to the $\gamma=0.1\kappa$ case, $\kappa$ is generally high and using a longer pulse $\kappa\tau\gg10$ is unlikely to present a bottleneck.}
    \label{fig:sine squared comparison}
\end{figure}

The formulae used for photon extraction probability in this paper are derived in the adiabatic limit, assuming an infinitely slow Raman drive pulse. This is algebraically straightforward and avoids the need to consider the details of the driving pulse while optimising the cavity parameters (once the cavity parameters are chosen, the drive can then be optimised according to experimental requirements). However, practical schemes must operate in finite time, and indeed reliable and \emph{fast} photon production underpins many of the target applications for this work. Using a finite-time pulse always reduces the output probability relative to the adiabatic result; the important consideration for the validity of our approach (that is, the separation of the cavity parameters from the driving pulse) is whether the ion-cavity system is likely to be driven rapidly enough to deviate significantly from the infinite-time result.

The validity of this assumption will vary depending upon many experimental factors including the parameters $g_0$, $\kappa$ and $\gamma$, and the characteristic timescale and exact pulse-shape of the driving pulse used. However, for typical smooth pulse profiles, the most important criterion for determining the output probability is the dimensionless quantity $\kappa \tau$, where $\tau$ is the width of the finite-time window in which the photon must be produced. To test this, we simulated the integrated output probability versus $\tau$ for a variety of $\Lambda$ type systems driven with a sine-squared pulse amplitude (as used in~\cite{Molony2016}), with the peak amplitude and pulse width chosen to maximise the output for each $\tau$.  $\Lambda$ systems were chosen with a range of cooperativities $C=\{0.1, 1, 10\}$ and cavity to atomic decay rate ratio $\kappa/\gamma=\{0.1,1,10\}$ in order to account for the variety of cavity parameters in current microcavity experiments.  The results, presented in Fig~\ref{fig:sine squared comparison} show that the output approximates the adiabatic limit for $\kappa \tau \gtrsim 10$. Therefore, as long as the photon production window is several times longer than the characteristic decay time of the cavity, the achievable output is very similar to the adiabatic limit and thus the latter provides an adequate approximation while optimising the cavity design.

The specific systems considered in this work cover cavity decay rates where ${20~\mathrm{MHz} \lesssim \kappa/2\pi\lesssim 300~\mathrm{MHz}}$, which will satisfy the adiabaticity condition for ${500~\mathrm{ns} \gtrsim \tau \gtrsim 33~\mathrm{ns}}$. These times are smaller than or comparable to other processes required during each photon production attempt, such as optical pumping for state preparation, optical modulator latency and other control system related overheads, and, in larger networks, the photon propagation time. We therefore consider that achieving near-adiabatic performance is possible for all the systems we consider, without a significant reduction in achievable repetition rate, although we note that for the lowest-$\kappa$ cavities we consider, these effects are certainly worthy of careful consideration if using repetition rates approaching \SI{1}{\mega\hertz}. In these cases, it should also be noted that the values presented here represent a lower bound on the output achievable for a pulse of this length, given that the pulse we use has a predefined shape and frequency; carefully designing the driving pulse can help increase the output probability~\cite{Vasilev2009} for lower $\kappa\tau$, though the improvement is typically modest over the functions used here.

\end{document}